       \def\oldlatex{0}
\ifnum\oldlatex=0
  \documentclass[11pt]{article}
  \usepackage[reqno]{amstex}
  \usepackage{epsfig}
  \newcommand{\e}{\ensuremath{\,\mathrm{e}}}
\else
  \documentstyle[11pt]{amsart}
  \newcommand{\e}{{\rm e}}
  \newcommand{\emph}[1]{{\em #1}}   % Is defined in LaTeX2e
  \newcommand{\mathcal}[1]{{\cal #1}}   % Is defined in LaTeX2e
  \newcommand{\text}[1]{{\mbox{#1}}}   % Is defined in LaTeX2e
  \newenvironment{align}[1]{\begin{equation}}%
  {\end{equation}}   % Is defined in LaTeX2e
  \newcommand{\displaybreak}{\linebreak}%
\fi

\newlength{\AmsBug}
\newcommand{\BugFix}{\hspace{\AmsBug}}

\begin{document}

%  \bibliographystyle{cqg}
%  Classical and Quantum Gravity Style

\addtocounter{page}{-1}

\title{
\raisebox{4cm}{{\small gr-qc/9608060}{\hspace{7cm}}{\small CERN-TH/96-92}}\\
{\mbox{ }}{\vspace{-3cm}}{\mbox{ }}\\
Anisotropic domain walls}

\author{%
Bj{\o}rn Jensen\\
{\em {\small Institute of Physics, University of Oslo, P.O. Box 1048
Blindern,}}\\
{\em {\small N-0316 Oslo, Norway}}
\and
Harald H. Soleng\thanks{Present address: 
NORDITA. Blegdamsvej 17, DK-2100 Copenhagen {\O}, Denmark}\\
{\em {\small Theory Division, CERN, CH-1211 Geneva 23, Switzerland}}}

\maketitle

\abstract{%
We find
an anisotropic, non-supersymmetric generalization of the extreme
supersymmetric domain walls of simple non-dilatonic supergravity
theory. As opposed to the isotropic non-\ and ultra-extreme domain
walls, the anisotropic non-extreme wall
has the \emph{same} spatial topology
as the extreme wall. The solution has naked singularities which
vanish in the extreme limit. Since the Hawking temperature on the
two sides is different,
the generic solution is
unstable to Hawking decay.}

\thispagestyle{empty}

{\mbox{ }}{\vspace{2cm}}\\
CERN-TH/96-92\\
August 1996

\newpage

\section{Introduction}

Domain walls
\cite{Vilenkin:PRep85}
are surfaces
interpolating between
regions with different expectation values of some matter
field(s).
Such objects are
interesting for a
variety of reasons.
Whenever the vacuum manifold has a non-trivial homotopy group
$\pi_0 (\mathcal{M})$, domain walls can exist
as topological defects \cite{Kibble:JPA76}.
Therefore, the possibility of domain wall formation in the
early universe---as a result of spontaneous symmetry breaking
in unified gauge theories---has attracted much interest.
But domain walls, and more generally solitons,
are of interest from a purely theoretical perspective.
Within string theory, it has recently been recognized
that Bogomol'nyi--Prasad--Sommerfield saturated states
could play an important r\^ole in
its non-perturbative dynamics.

Over the last few years domain walls have been studied
within
four-dimensional $N=1$ supergravity theory (see
Ref.~\cite{CS:Prep96} for a review).
After discovery of the ``ordinary'' supersymmetric
supergravity domain walls \cite{CGR:NPB92,CG:PLB92},
their global space--time structure has been
analyzed \cite{CDGS:PRL93,Gibbons:NPB93},
and the relation to the
corresponding non-supersymmetric
domain wall bubbles has been
clarified \cite{CGS:PRL93,CGS:PRD93} (see
Refs.~\cite{Cvetic:PRL93,Cvetic:PLB94,CS:PRD95} for generalizations to
the dilatonic case).
The isotropic vacuum domain walls can be
classified according to the value of their surface
density $\sigma$, compared to the energy-densities
of the vacua outside the wall \cite{CGS:PRL93,CGS:PRD93}.
The three kinds of isotropic walls are
the static planar
\emph{extreme walls} with
$\sigma=\sigma_{\text{ext}}$,
the
\emph{non-extreme} two-centred bubbles
with
$\sigma>\sigma_{\text{ext}}$,
and the \emph{ultra-extreme}
vacuum decay bubbles
with
$\sigma<\sigma_{\text{ext}}$.

In this paper we consider the \emph{anisotropic} case,
where the various components of the
metric tensor has a different functional
dependence of the distance $z$ from the wall.
We shall restrict the analysis to
space--times with a line
element of the
form\footnote{We use units so that $\kappa\equiv 8\pi G=c=1$.}
\begin{equation}
ds^2= A_{(t)}^2 dt^2
    - A_{(x)}^2 dx^2 - A_{(y)}^2 dy^2 - dz^2
\label{Eq:Metric}
\end{equation}
where $A_{(i)}$ ($i\in \{t,x,y\}$) all are functions of $z$, and
where we have used the gravitational
gauge (coordinate) freedom to
normalize $g_{zz}$ to $-1$.
Throughout this paper, indices in parentheses, such as those in
the metric above, \emph{shall not} be subject to the Einstein summation
convention.

\section{The supersymmetric solution}

Consider the bosonic piece of an $N=1$ supersymmetric theory
with one chiral matter superfield $\mathcal{T}$
in $3+1$ space--time dimensions:
\begin{align}
\label{Eq:Lagrangian}
\mathcal{L}&=-\frac{1}{2} R + K_{T\overline{T}} \partial^\mu \overline{T}
\partial_\mu T
-V(T,\overline{T})
\tag{\ref{Eq:Lagrangian}a}
\displaybreak[3]\\
\intertext{\BugFix
where $K(T,\overline{T})$ is the K\"ahler potential,}
V(T,\overline{T})&= \e^K \left( K^{T\overline{T}} |D_T W|^2-3 |W|^2
\right),
\tag{\ref{Eq:Lagrangian}b}
\displaybreak[3]\\
\intertext{\BugFix
is the scalar potential, and}
D_T W &\equiv \e^{-K} \left[ \partial_T (\e^K W)\right] =
W_T + W K_T
\tag{\ref{Eq:Lagrangian}c}%
\end{align}\addtocounter{equation}{1}%
is the K\"ahler covariant derivative acting
on the superpotential $W$.

In a supersymmetric vacuum $D_T W=0$, and thus the
effective cosmological constant in such a vacuum is
\[
\Lambda_{\text{susy}}= -3 \e^K |W|^2 \equiv -3\alpha^2.
\]
Hence,
it is non-positive in this theory.

\begin{figure}[htb]
\centering{\epsfig{%
file=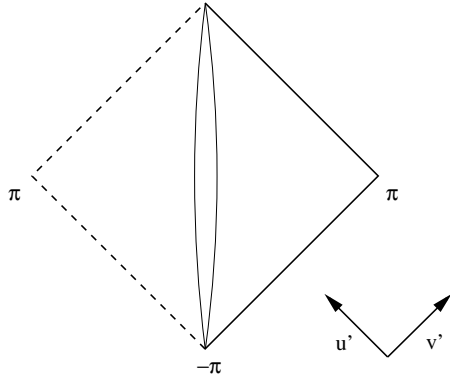,% 
height=5cm,%
clip=}}
\caption{Penrose conformal diagram of one diamond formed by
compactifying the coordinates $(t,z)$. The wall is the
lens-shaped region splitting the diamond in half. To the
right of the wall is the Minkowski region. Solid lines
symbolize geodesically completeness. To the left is the AdS
region, where dashed lines indicates the need for coordinate
extensions.}
\label{Fig:PenroseSUSYAdSM}
\end{figure}

Let us now review the supersymmetric 
anti-de~Sitter (AdS)--Min\-kow\-ski wall already
discussed in Refs.~\cite{CDGS:PRL93,CGS:PRD93,Griffies:PHD}.
In that case the metric (\ref{Eq:Metric}) takes the form
\begin{equation}
ds^2= A^2 (dt^2
    - dx^2 - dy^2) - dz^2
\label{Eq:IsoMetric}
\end{equation}
(cf.\ the Appendix). 
The behaviour of the function $A(z)$ some distance from
the wall, which is placed at $z = 0$, is given by
\begin{equation}
A(z) = \left\{
\begin{array}{ll}
\e^{\alpha z} & \text{for $z<0$: the AdS vacuum}\\
1            & \text{for $z>0$: the Minkowski vacuum}.
\end{array}\right.
\label{Eq:IsoA}
\end{equation}
The fact that $A(z)$ vanish as $z \rightarrow -\infty$
suggests that the line element is geodesically incomplete
on the AdS side of the wall. Further investigation 
shows that null geodesics leave the AdS side with finite
affine parameter. Since $A(z)$ is a function of $z$ only, the
(2+1)-dimensional space--times with constant $z$ 
(the slices parallel to the wall) are
simply Minkowski space. Therefore, the interesting
directions for possible coordinate extensions are $(t,z)$.
A Penrose conformal diagram for the compactified
$(t,z)$ coordinates is shown in Fig. \ref{Fig:PenroseSUSYAdSM}.

\section{Anisotropic vacuum solutions}

In the thin wall approximation the energy--momentum tensor
of the wall-forming
matter field(s) is approximated by a cosmological constant 
outside the wall where the (nearly constant) potential term is dominating 
and a delta-function singularity in the wall surface where the
kinetic term is dominating.

Let us first look at the gravitational field off the wall.   
Using the natural orthonormal frame, and the definitions
\begin{equation}
H_{(i)} \equiv \frac{d\ln A_{(i)} (z)}{dz},
\label{Eq:H_def}
\end{equation}
the Einstein tensor for the metric (\ref{Eq:Metric}) is
\begin{align}
\label{Eq:Einstein}
{G^0}_0   &= -H_{(x)}^2 - H_{(x)} H_{(y)} - H_{(y)}^2
- H_{(x)}' - H_{(y)}'
\tag{\ref{Eq:Einstein}a}
\displaybreak[3]\\
{G^1}_{1} &= -H_{(t)}^2 - H_{(t)} H_{(y)} - H_{(y)}^2 - H_{(t)}' - H_{(y)}'
\tag{\ref{Eq:Einstein}b}
\displaybreak[3]\\
{G^2}_{2} &= -H_{(t)}^2 - H_{(t)} H_{(x)} - H_{(x)}^2 - H_{(t)}' - H_{(x)}'
\tag{\ref{Eq:Einstein}c}
\displaybreak[3]\\
{G^3}_{3} &= -H_{(t)} H_{(x)} - H_{(t)} H_{(y)} - H_{(x)} H_{(y)}
\tag{\ref{Eq:Einstein}d}
\end{align}\addtocounter{equation}{1}%
where a prime means the derivative with respect to $z$.

We now solve the Einstein equations with the
stress-energy tensor of a vacuum
with a non-positive energy density
\[
{G^\mu}_\nu = {\delta^\mu}_\nu\Lambda, 
\]
where
$\Lambda$ is a negative cosmological constant
which we shall parametrize by $\Lambda = -3\alpha^2$.

By combining the Einstein equations in various ways, we get
\begin{align}
\label{Eq:Hdiffeqs}
& H_{(i)}' + 3H H_{(i)} = 3\alpha^2
\tag{\ref{Eq:Hdiffeqs}a}
\label{Eq:Hidiff}\\
& H' + 3H^2 = 3\alpha^2.
\tag{\ref{Eq:Hdiffeqs}b}
\label{Eq:Hdiff}
\end{align}\addtocounter{equation}{1}%
Here $H = \frac{1}{3} \sum_i H_{(i)}$.
Integrating Eq.~(\ref{Eq:Hdiff}), we get
\begin{equation}
H=\alpha \frac{ (\xi+\alpha)^2 \e^{6\alpha z} + \xi^2}
                 {(\xi+\alpha)^2 \e^{6\alpha z}-\xi^2}
\label{Eq:HSolutionLambda}
\end{equation}
where $\xi$ is an integration constant.

Given $H$, the solution for $H_i$ is easily found to be%
\footnote{The solution is related to a corresponding
generalization
\cite{Gron:PRD85b}
of the Kasner \cite{Kasner:AJM21}
cosmological solution
by a complex coordinate transformation, see Ref.~\cite{CS:PRD95}
for applications of such transformations
to the dilatonic domain walls.}
\begin{equation}
H_{(i)}= H + c_{(i)} h
\label{Eq:HiSolution}
\end{equation}
where
\begin{equation}
h= \frac{2 \alpha \xi (\alpha+\xi) \e^{3 \alpha z}}
        {(\alpha+\xi)^2 \e^{6\alpha z}-\xi^2}.
\end{equation}
Note that under the 
transformation
\begin{equation}
\xi\rightarrow \xi'=-\alpha\xi/(\alpha+2\xi), 
\label{Eq:transformation}
\end{equation}
$H$ is invariant and $h$ simply changes sign.

Moreover, 
both
$H$ and $h$ are invariant 
under the transformations 
\begin{equation}
\alpha\rightarrow \alpha'=-\alpha\ \ {\text{and}}\ \
\xi\rightarrow \xi'=\xi+\alpha. 
\label{Eq:sign_alpha_transformation} 
\end{equation}
The space--time geometry and 
the surface energy
of the wall
are therefore left unchanged under this transformation.
It is thus sufficient to study the case $\alpha>0$.

The constants satisfy
\begin{equation}
\sum_i c_{(i)}= 0\quad \text{and} \quad    \sum_i c_{(i)}^2 = 6,
\label{Eq:c_restrictions}
\end{equation}
which represents (in three
 dimensions) a plane going through origo
with normal vector making equal angles with all
three axes, and a spherical shell with
radius $\sqrt 6$ centred in origo respectively.
The allowed values for the constants therefore lie on
the circle where the plane cuts through the sphere, and it
is easy to verify that $|c_{(i)}| \leq 2$.
We note
that $\xi$ may be interpreted as an
anisotropy-parameter because one gets the isotropic
solution by letting $\xi\rightarrow 0$ or
$\xi\rightarrow -\alpha$.

By taking the limit $\alpha\rightarrow 0$,
we obtain the
Kasner \cite{Kasner:AJM21} type solution, where
$H_{(i)}$ is defined by taking
\begin{equation}
H_{\alpha=0} = h_{\alpha=0} = \frac{\xi}{1 + 3\xi z}
\label{Eq:Kasner}
\end{equation}
in Eq.~(\ref{Eq:HiSolution}).
Now we get the isotropic Minkowski solution
by letting $\xi\rightarrow 0$.

\section{Domain wall solutions}

We shall now match the above vacuum solutions
by means of an infinitely thin domain wall
junction. To this end we employ the Israel formalism \cite{Israel:INC66}
for singular
hypersurfaces and thin shells. 

\subsection{Israel matching}

We place the wall at $z=0$,
and define the spacelike unit normal vector to its surface by
\begin{equation}
\mbox{\boldmath $n$} \equiv \frac{\partial}{\partial z}\ \ \text{and}
\ \ \mbox{\boldmath $n \cdot n$} = n^\mu n_\mu \equiv -1.
\end{equation}
The extrinsic curvature
of the wall is a three-dimensional tensor whose components
are defined by the covariant derivative of
this unit normal:
\begin{equation}
{K^i}_j \equiv -{n^i}_{;j}.
\end{equation}
For our choice of coordinates we have
\[
K_{ij} = -\zeta \frac{1}{2}g_{ij,z},
\]
where $\zeta=\pm 1$ is a sign factor depending on the
direction of the unit normal. For a kink-like matter source
$\left. \zeta\right|_{-}=
\left. \zeta\right|_{+}=1$, so
in the sequel we drop this sign factor.
Hence, we get
\begin{equation}
{K^i}_{j} = -{\delta^i}_{j} H_{(i)}.
\label{Eq:Kuilj}
\end{equation}
The stress-energy tensor for
the wall is given by
\[
{S^i}_{j} = -({\gamma^i}_{j} - {\delta^i}_j 
{\gamma^k}_k),
\]
where ${\gamma^i}_j \equiv \lim_{\epsilon \rightarrow 0}
[{K^i}_j%
(z = +\epsilon) - {K^i}_j(z = -\epsilon)]$.
Using
Eq.~(\ref{Eq:Kuilj}) and
inserting the
solution~(\ref{Eq:HiSolution}),
we get
\begin{equation}
{S^i}_j  = -{\delta^i}_j\left[
c_{(i)} h
+2H
\right]_{z=-\epsilon}^{z=+\epsilon}.
\end{equation}
The square brackets stand for the \emph{difference} taken
at the points indicated with the super-\ and subscript
on the closing bracket.
Now, for a domain wall we must have
a surface energy density
$\sigma\equiv {S^t}_t$ and a tension $\tau\equiv{S^x}_x={S^y}_y=\sigma$,
\emph{i.e.}, a boost invariant energy--momentum
tensor
on the world volume.
Together with Eqs.~(\ref{Eq:c_restrictions})
this implies that
\begin{equation}
\left.  c_{(i)}\right|_{-}=\left. \lambda\, c_{(i)} \right|_{+}
\ \ \text{and}\ \
\left. h \right|_{-}=\left. \lambda\, h \right|_{+},
\label{Eq:plusminus}
\end{equation}
where
$\lambda=\pm 1$.

\subsection{Wall between two vacua with $\mathbf{\Lambda=0}$}

In this case the vacuum solutions on both sides of the wall
are given by Eqs.~(\ref{Eq:Kasner}).
The domain wall equation of state $\sigma=\tau$, together with the
constraints on the $c_{(i)}$-constants
gives the following expression when we
choose $\Lambda = 0$ on
both sides of the wall:
\begin{equation}
\sigma= 4  \xi_{-}
\end{equation}
where
the parameters are related as
\begin{equation}
\xi_{-} = -\xi_{+}\ \
\text{and}\ \
\left. c_{(i)} \right|_{-} = -\left. c_{(i)}\right|_{+}
\end{equation}
or $\lambda=-1$.
(With $\lambda=1$ and
$\xi_{-} = \xi_{+}$
there is no wall at all.)
Thus, in order to have a domain wall
with a positive energy density,
we must have $\xi_{-}>0$.
We note that
$H>0$ for $z<0$ and $H<0$ for $z>0$.
Hence, the average scale factor
is decreasing away from the wall
on both sides.
This is a non-extreme solution
for which the extreme limit is trivial Minkowski space--time.
It was discovered by Tomita \cite{Tomita:PLB85}.

\subsection{Walls between  $\mathbf{\Lambda<0}$ and
$\mathbf{\Lambda=0}$ vacua}

Let $\Lambda<0$ for $z<0$ and
$\Lambda = 0$ for $z>0$.
Then
\begin{equation}
\sigma  = 2\alpha_{-} +4 \xi_{-},
\end{equation}
where again $\lambda=-1$
relates the values of $h$ and $c_{(i)}$ on
each side of the wall as in Eq.~(\ref{Eq:plusminus}). Selecting
$\lambda=1$ adds nothing new due to the invariance of $H$ and $h$
under the transformation given in Eq.~(\ref{Eq:transformation}).
For the wall to have a positive energy density, 
$\alpha_{-}>-2\xi_{-}$.
Now the anisotropy parameter on the side with vanishing
cosmological constant is related
to the parameters on the other side by 
\begin{equation}
\xi_{+}= \frac{-2\xi_{-} (\alpha_{-} +\xi_{-})}
             {\alpha_{-} + 2 \xi_{-}}.
\end{equation}
In the isotropic limit $\xi\rightarrow 0$, we
recover the extreme
anti-de~Sitter--Minkowski wall \cite{CGR:NPB92,CG:PLB92,CDGS:PRL93}.

\subsubsection{Non-extreme solution}
\label{subsub:Smooth_Non-extreme}

If $\alpha_{-}, \xi_{-} > 0$,
then the solution is
a kink-like
non-extreme solution with $\sigma>\sigma_{\text{ext}}=2\alpha$.
It is smoothly related to the extreme solution in the limit
$\xi\rightarrow 0$.

\subsubsection{Ultra-extreme solution}

If $\alpha_{-} >0 $ and $-\alpha_{-}/2 < \xi_{-} < 0$,
then the solution is
a kink-like
ultra-extreme solution with $\sigma < \sigma_{\text{ext}}=2\alpha$.
The average scale factor is monotoneously increasing from
the $\Lambda<0$ vacuum through the wall and into the
$\Lambda=0$ vacuum.
This solution is smoothly related to the extreme solution in the limit
$\xi\rightarrow 0$.

Physically this solution would correspond to a planar vacuum decay wall,
but since the Euclidean
action would be infinite in this case, only $\mathrm{O}(4)$
symmetric bubbles are expected to be realized by vacuum tunnelling.
We therefore regard this solution as unphysical.

\subsection{Space--time structure and Hawking temperature}

\begin{figure}[htb]
\centering{\epsfig{%
file=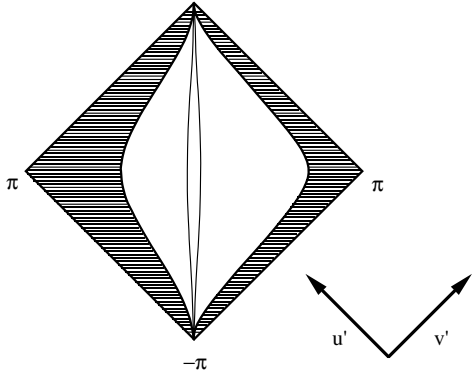,%
height=5cm,%
clip=}}
\caption{Penrose conformal diagram 
formed by compactifying the coordinates $(t,z)$. 
Compare this diagram with Fig.~\protect\ref{Fig:PenroseSUSYAdSM}. 
The wall is again the lens-shaped region splitting the diamond in half. 
To the right of the wall is the Kasner region, and to the left is the AdS
region. The shaded area on both sides represents 
the forbidden regions beyond 
the singularities.  
They are \emph{not} part of the classical space--time.}
\label{Fig:PenroseAdSK}
\end{figure}

In Fig.~\ref{Fig:PenroseAdSK} we present a Penrose conformal diagram 
for the compactified $(t,z)$ coordinates of the space--time discussed 
in Sect.~\ref{subsub:Smooth_Non-extreme}. 
On both sides of the wall
there are naked singularities.

In general, 
the naked singularities have infinite 
Hawking temperatures.
Here we calculate the temperature (equivalently, the surface gravity)
at a wall-induced singularity. 
It is given by
\begin{equation}
T =  -\frac{ z_{\text{sng}} }{ |z_{\text{sng}} |} 
\frac{\hbar}{2\pi k} 
\lim_{z \rightarrow z_{\text{sng}}} A_{(t)}',
\label{HawkingT}
\end{equation}
where $k$ is Boltzmann's constant, and 
\begin{equation}
z_{\text{sng}} = 
\left\{ \begin{array}{ll}
\frac{1}{6\alpha}\ln\frac{\xi^2}{(\alpha + \xi)^2} &
{\text{if }} \alpha\neq 0\\
-\frac{1}{3\xi} & {\text{if }} \alpha=0
        \end{array}
\right.
\label{Eq:zsng}
\end{equation}
is the value of 
$z$ at the singularity. Inserting for $A_{(t)}'$ and 
rearranging, we get a dimensionless and finite representation of the 
temperature:
\begin{equation}
T\frac{2\pi k}{\hbar \xi L} = -\frac{ z_{\text{sng}} }{ |z_{\text{sng}}
|}(1 + c_{(t)}) 
\left[\frac{2\xi}{\alpha + 2\xi}\right]^{(1 - c_{(t)})/3} 
\left[\frac{\alpha + \xi}{\xi}\right]^{1/3}, 
\label{T../L}
\end{equation}
where
\begin{equation}
L \equiv 
\lim_{z \rightarrow z_{\text{sng}}} 
\left\{ 
\begin{array}{ll}
\left[ \frac{(\alpha + 
\xi) \e^{3\alpha z} - \xi}{\alpha} \right]^{(c_{(t)} - 2)/3}
& {\text{if }} \alpha\neq 0\\
\left( 
1+3\xi z \right)^{(c_{(t)} - 2)/3}
& {\text{if }} \alpha= 0
\end{array}
\right.
\label{L}
\end{equation}
with the appropriate 
$z_{\text{sng}}$ as defined in 
Eq.~(\ref{Eq:zsng}). 
It is understood that 
$\xi_{\pm}$ and $\left. c_{(t)}\right|_{\pm}$
are used in place of $\xi$ and $c_{(t)}$ in the above
expressions.

\begin{figure}[htb]
\centering{\epsfig{%
file=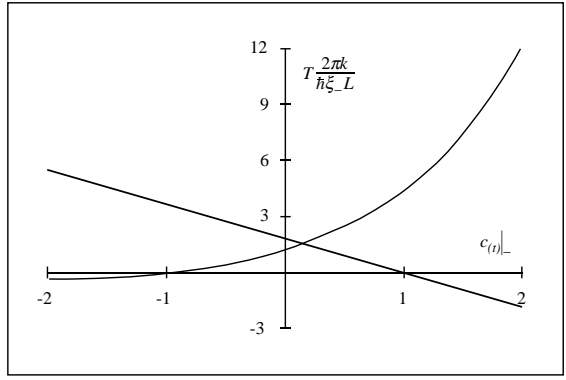,%
height=5cm,%
clip=}}
\caption{Plot of $T\frac{2\pi k}{\hbar \xi_{-}
L}$ versus $\left.  
c_{(t)}\right|_{-}$ for the case discussed in
Sect.~\protect\ref{subsub:Smooth_Non-extreme}
with $\xi_{-} =
\alpha_{-}/10$. 
The curved graph represents the temperature on the $\Lambda <
0$, $z < 0$ 
side of the wall, while the straight graph yields the temperature
on the $\Lambda = 
0$, $z > 0$ side of the wall. Bear in mind that these
graphs must be 
multiplied with a factor $L$, which is generally
diverging, in order 
to obtain the temperature. Note that 
$|c_{(t)}|\leq 1$ 
in order to have a non-negative 
temperature everywhere.} 
\label{Fig:temp}
\end{figure}

At the singularities
$L$ diverges for all $c_{(t)}$ except $c_{(t)} = 2$, for which it 
equals $1$. A plot of $T\frac{2\pi k}{\hbar \xi_{-} L}$ versus $\left. 
c_{(t)}\right|_{-}$ 
for the case discussed in Sect.~\ref{subsub:Smooth_Non-extreme}
is provided in Fig.~\ref{Fig:temp}.

In order to have non-negative temperatures on both sides of
the wall $|c_{(t)}|\leq 1$. However, in general the 
Hawking temperatures are different on the two sides,
and therefore we expect the anisotropic domain wall solutions to
be unstable to Hawking decay.

\section{Conclusion}

Only the planar, static and isotropic 
domain wall has a Killing spinor. 
Gravitational anisotropy 
therefore breaks supersymmetry and the corresponding
non-extreme topological defects have 
a larger surface energy density. 

The ultra-extreme walls 
are planar vacuum decay walls with a smaller energy-density. 
Due to their infinite Euclidean action, we consider 
these planar ultra-extreme solutions
to
be unphysical.

The anisotropic domain walls 
generally
have a non-vanishing temperature gradient. These solutions are 
therefore unstable to Hawking decay.

\appendix

\section{Killing spinor implies isotropy}

The supercovariant derivative acting on
the Majorana 4-spinor $\epsilon$ is given by
\[
\widehat{\nabla}_\rho \epsilon=
\left[
2\nabla_\rho + i\e^{K/2}(W P_{\text{R}}
+ \overline{W}P_{\text{L}})\gamma_\rho - \text{Im}(K_T\partial_\rho
T)\gamma^5
\right]
\epsilon.
\]
With a static and anisotropic line element of
the planar form (\ref{Eq:Metric}),
we get the following explicit form of the
supercovariant derivative acting on the spinor
\begin{align*}
\widehat{\nabla}_t \epsilon &= \left[2\partial_t +
\partial_z A_t\gamma^0\gamma^3 + iA_t\gamma^0(W P_{\text{L}}
+ \overline{W} P_{\text{R}}) \e^{K/2}\right]\epsilon\\
\widehat{\nabla}_x \epsilon &= \left[2\partial_x
+ \partial_z A_x\gamma^1\gamma^3
- i A_x\gamma^1(W P_{\text{L}} + \overline{W} P_{\text{R}})
\e^{K/2}\right]\epsilon\\
\widehat{\nabla}_y \epsilon &= \left[2\partial_y
+ \partial_z A_y\gamma^2\gamma^3 - i A_y\gamma^2(W P_{\text{L}} +
\overline{W}
P_{\text{R}})\e^{K/2}\right]\epsilon\\
\widehat{\nabla}_z \epsilon &= \left[2\partial_z
- i\gamma^3(W P_{\text{L}} + \overline{W} P_{\text{R}}) \e^{K/2}
- \gamma^5\text{Im}(K_T\partial_z T)\right]\epsilon.
\end{align*}
The Killing spinor is one which satisfies the equation
$\widehat{\nabla}_\rho \epsilon = 0$. Using the Weyl basis
\[
\begin{array}{ccc}
\gamma^0 = \left( \begin{array}{cc} 0 & 1 \\ 1 & 0 \end{array} \right) &
\gamma^i = \left( \begin{array}{cc} 0 & \sigma^i \\ -\sigma^i & 0
\end{array} \right) & \gamma^5 = \left( \begin{array}{cc} -i & 0 \\ 0 & i
\end{array} \right),
\end{array}
\]
where $\sigma^i$ are the Pauli matrices,
the Majorana spinor $\epsilon = \epsilon^c$ is
\begin{equation}
\epsilon = \left(\begin{array}{c}
                  \epsilon_1 \\
                  \epsilon_2 \\
                  \epsilon_2^* \\
                 -\epsilon_1^*
                 \end{array} \right).
\end{equation}
Additionally, there is a constraint on the
supersymmetric parameter $\epsilon_1 = \e^{i\Theta}\epsilon_2^*$.
A Killing spinor therefore calls
for the following equations to be satisfied
\begin{equation}
\partial_z \ln A_{(i)} = H_{(i)} = iW\e^{K/2}\e^{-i\Theta}
\label{Eq:KillingEq}
\end{equation}
which implies $H_{(i)}=H$, satisfied only
in the isotropic limit $\xi \rightarrow 0$.

%%%%%%%%%%%%%%%%%%   REFERENCES   %%%%%%%%%%%%%%%%%%%%%%%%
%

\end{document}